\documentclass[prb,twocolumn,showpacs]{revtex4}

\usepackage{graphicx}
\usepackage{amsmath}
\usepackage{mathptmx}

\newcommand{\LC}{La$_{2/3}$Ca$_{1/3}$MnO$_{3}$ }

\begin{document}

\title{Large and Small Polaron Excitations in
La$_\mathbf{2/3}$(Sr/Ca)$_\mathbf{1/3}$MnO$_\mathbf{3}$ Films}

\author{Ch.~Hartinger}
\affiliation{EP V, Center for Electronic Correlations and Magnetism, Augsburg
University, 86159 Augsburg, Germany}

\author{F.~Mayr}
\affiliation{EP V, Center for Electronic Correlations and Magnetism, Augsburg
University, 86159 Augsburg, Germany}

\author{J.~Deisenhofer}
\affiliation{EP V, Center for Electronic Correlations and Magnetism, Augsburg
University, 86159 Augsburg, Germany}

\author{A.~Loidl}
\affiliation{EP V, Center for Electronic Correlations and Magnetism, Augsburg
University, 86159 Augsburg, Germany}

\author{T.~Kopp}
\affiliation{EP VI, Center for Electronic Correlations and Magnetism, Augsburg
University, 86159 Augsburg, Germany}

\date{\today}

\begin{abstract}
We present detailed optical measurements of the mid-infrared (MIR) excitations
in thin films of La$_{2/3}$Sr$_{1/3}$MnO$_{3}$ (LSMO) and
La$_{2/3}$Ca$_{1/3}$MnO$_{3}$ (LCMO) across the magnetic transition. The shape
of the excitation at about 0.2~eV in both samples is analyzed in terms of
polaron models. We propose to identify the MIR resonance in LSMO as the
excitation of large polarons and that in LCMO as a small polaron excitation. A
scaling behavior for the low-energy side of the polaronic MIR resonance in LSMO
is established.
\end{abstract}


\maketitle

Historically, polarons were best identified in measurements with
charge carriers in non-metals (e.~g., from F-centers in alkali
halides) or doped semiconductors. More recently, the concept of
polaronic excitations has again been in the focus of solid state
research with the advent of strongly correlated electronic
systems, especially with the discoveries of high-temperature
superconductivity and the colossal magnetoresistance in thin films
of \LC (LCMO) \cite{Helmolt93}. In both classes of these
correlated materials, cuprates and manganites, it is now widely
accepted that, in the presence of a strong electron-phonon
coupling, polaronic effects have to be considered an important
ingredient to understand the complex physical properties of these
compounds \cite{Calvani01}.

The fingerprints of polarons in the manganites are usually
associated with the temperature dependence of high-temperature
dc-resistivity \cite{Worledge96} and the occurrence of a
mid-infrared (MIR) excitation in the optical conductivity ---
termed polaron peak in the following --- which has been observed
by several authors in both single crystalline
\cite{Okimoto97,Jung98,Kim98,Saitoh99,Lee99} and thin film
manganites \cite{Kaplan96,Quijada98,Machida98}. Though evidence of
the signature of polarons in the optical conductivity
\cite{Reik67,Goovaerts73,Emin75} has been reported abundantly, the
question whether the charge carriers are rather delocalized (large
polarons, LP) or strongly localized in a locally polarized lattice
(small polarons, SP) remains still under debate in the case of
manganites.


Theoretically,  SP and LP have been investigated in several
studies \cite{Roeder96,Lee97,Millis96} on doped manganites, but
the observed polaron peak has been evaluated mainly in terms of
the SP model or through Gaussian or Lorentzian fits for a
phenomenological description \cite{Jung98,Kim02}. Kim
\textit{et~al.}~reported on experimental evidence for a LP
excitation in polycrystalline LCMO \cite{Kim98}, but unfortunately
they did not identify the LP through a fit to the proposed LP
model by Emin \cite{Emin93}.

The purpose of this paper is to elucidate the nature of the
polaronic charge carriers by comparison of optical spectroscopy
measurements in thin films of LCMO and
La$_{2/3}$Sr$_{1/3}$MnO$_{3}$ (LSMO). In contrast to single
crystals the use of thin films has the advantage of a lower
conductivity due to grain boundaries and internal strain.
Consequently, screening is effectively reduced
--- to a level where phonons and polarons are well observable in the optical
conductivity, even in the metallic phase. This allows the detailed analysis of
polarons from the respective optical data.

\begin{figure}[htb]
\centering
\includegraphics[width=.45\textwidth,clip]{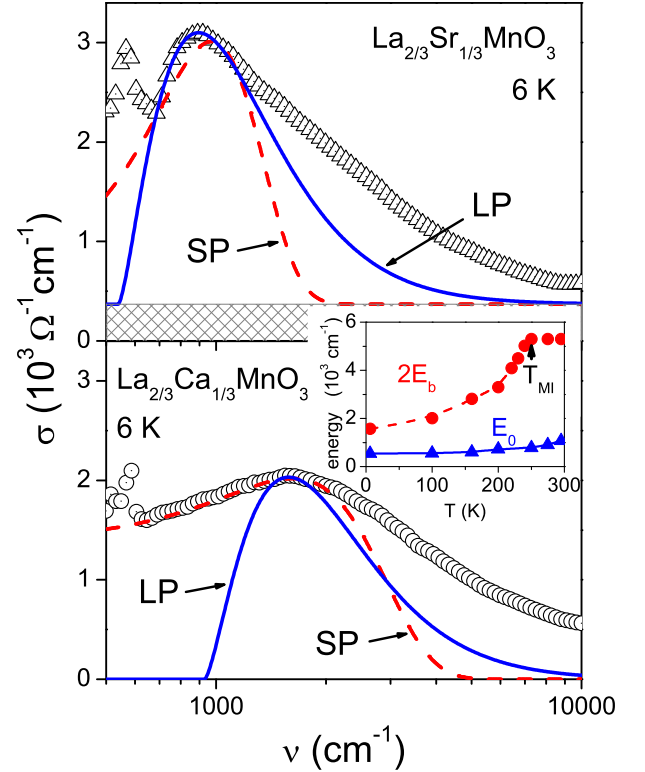}
\caption[]{\label{fit}Comparison between fit and experiment for
the MIR excitation of LSMO (upper panel) and LCMO (lower panel) at
6~K using a LP model (Eq.~(\ref{eq:EminPolaron}),solid line) and a
SP model (Eq.~(\ref{eq:polaron}),dashed line). Inset: polaron
binding energy for LCMO ($2 E_{b}$) and for LSMO ($E_{0}$).}
\end{figure}

We demonstrate that the distinctive shape of the polaron peak in
LCMO differs significantly from the one in LSMO allowing to
distinguish between SP in LCMO and LP in LSMO. A striking
characteristic of the polaron peak in LSMO is the observed scaling
behavior for the low-energy side (see Fig.~\ref{fit}). We will
confirm this scaling in a weak coupling evaluation, analogous to
that of Tempere and Devreese for \textit{an interacting
many-polaron gas} \cite{Tempere01}.

As the lattice distortion is of major importance for the effect of charge
carrier localization, the structural differences between LCMO, which is
orthorhombically distorted, and LSMO, which reveals a rhombohedral symmetry,
represent promising conditions to find distinct regimes of the electron-phonon
coupling strength $\alpha$ and, consequently, to distinguish between competing polaronic
models.

Thin films have been prepared using a standard pulsed laser deposition
technique \cite{Christey94}. They were grown onto single crystalline
substrates: LSMO on (LaAlO$_{3}$)$_{0.3}$(Sr$_{2}$AlTaO$_{5}$)$_{0.7}$ and LCMO
on NdGaO$_{3}$. The typical thin film thickness was between 200~nm and 400~nm.
X-ray analysis revealed a rhombohedral structure for LSMO and a preferred
growth along the [100] axis. For LCMO an orthorhombic structure was found and
the growth direction was [110]. Below room temperature the reflectivities of
the sample and the pure substrate were measured using the Fourier transform
spectrometers Bruker IFS 113v and IFS 66v/S, to cover the frequency range from
50 to 40000~cm$^{-1}$. In addition, the reflectance for frequencies from 10 to
30~cm$^{-1}$ was calculated from the measured complex conductivity data, which
were obtained by submillimeter-transmission measurements using a Mach-Zehnder
type interferometer. This set up allows both, the measurement of the
transmittance and the phase shift of a film on a substrate. The combined data
sets were used for a Kramers-Kronig analysis to obtain the optical conductivity
$\sigma$, the real part of the complex conductivity.

\begin{figure}[h]
\centering
\includegraphics[width=.5\textwidth,clip,angle=0]{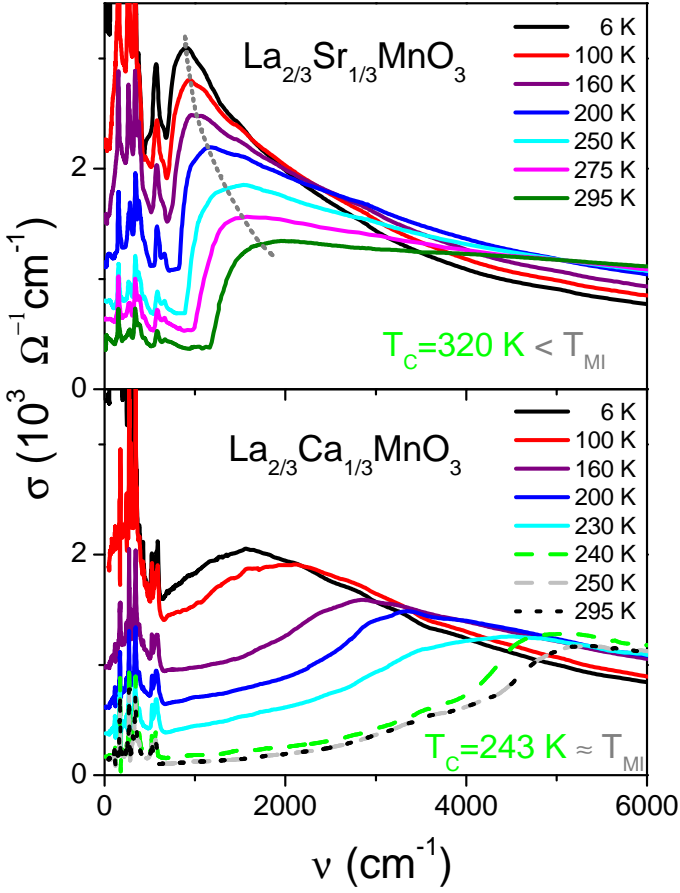}
\caption[]{\label{sigma}Optical conductivity $\sigma$ for
6~K~$\leq T\leq$~295~K. Straight lines represent data in the FM,
dashed and dotted lines indicates the PM phase. Upper panel: LSMO
($d=400$~nm). The dotted line is to guide the eyes for the shift
of the maximum of the MIR excitation. Lower panel: LCMO
($d=200$~nm).}
\end{figure}

In Fig.~\ref{sigma} we present $\sigma(\nu)$ for LSMO (upper
panel) and LCMO (lower panel) at different temperatures. In the
case of LCMO, with a Curie-temperature $T_{C}=243$~K, the spectra
in the complete frequency range were measured both in the
paramagnetic (PM) and in the ferromagnetic (FM) regime. The data
for LSMO, as shown in the upper panel of Fig.~\ref{sigma}, only
cover the FM regime. We defined $T_{C}$ as the inflection point
and $T_{\rm MI}$ as the point of the maximum in the dc-resistivity
curve.

In LCMO an almost symmetric but strongly temperature dependent
maximum is observed in the MIR range which suffers a continuous
loss of the spectral weight upon increasing the temperature up to
$T_{C}$. Above $T_{C}$ the temperature dependence is negligible.
The maximum moves from 1800 to 5300~cm$^{-1}$. The loss in
spectral weight can be explained in terms of a decreasing carrier
mobility \cite{Jakob98}. These findings are in agreement with
previous studies of Ca-doped manganites \cite{Kim98,Quijada98}.
For LSMO, in contrast, there is a sharp onset which, along with
the asymmetric peak, shifts by approximately a factor of 2 in
energy in the range $6\,\mathrm{K}<T<295\,\mathrm{K}$. This trend
continues up to the insulating phase, where the reflectivity data
(not shown) do not change any further \cite{Hartinger03}.


The prominent feature of the optical spectra, on which we will
concentrate in the following, is the distinctly different shape of
the polaron peak for LSMO and LCMO. In LSMO the excitation is
clearly asymmetric with a steep rise below and a long tail above
the peak position, whereas for LCMO a broader hump is visible
which is more symmetric about its maximum (compare
Fig.~\ref{sigma}). A fit with standard single-polaron models
highlights the remarkable qualitative differences and allows a
preliminary identification (see the respective optical spectra at
6~K in Fig.~\ref{fit}). For the LP fit we took the
phenomenological approach by Emin \cite{Emin93} which assumes a
photoionization of the charge carriers from self-trapped into free
carrier states. The threshold behavior of the respective optical
conductivity
\begin{equation}
\label{eq:EminPolaron} \sigma(\omega)\;=\; n_p \frac{64}{3}
\frac{e^2}{m}\;\frac{1}{\omega}\;
\frac{\bigl(k(\omega)\,R\bigr)^3}{\bigl[1+(k\left(\omega)\,R\right)^2 \bigr]^4}
\end{equation}
reproduces the observed upturn at the low-frequency side of the
polaron peak. Here,
$k(\omega)=\sqrt{2m\hbar(\omega-\omega_0)}/\hbar$, with the
temperature-dependent threshold frequency for the absorption
$\omega_0=E_0/\hbar=2\pi \nu_0$ (see inset of Fig.~\ref{fit}).
$E_0$ is interpreted as the energy difference between the
localized ground state and the lowest continuum state. The radius
of the hydrogenic ground state of the polaron is $R$, the density
of polarons is $n_p$, and $m$ is the (effective) mass of the free
carrier states at the band minimum. For $m=3m_e$ ($m_e$ is the
free electron mass) and $\nu_0 =540$~cm$^{-1}$, we find $R/a=0.8$
with the lattice constant $a= 3.88$~\AA\ and $n_p\simeq 4 \times
10^{21}$~cm$^{-3}$ at 6 K. We challenge this phenomenological
approach below and propose a microscopic modelling.

The SP is interpreted in terms of the Holstein small-polaron
theory. It has been proposed to apply an extension of the standard
SP optical conductivity formula \cite{Reik72,Emin75} to the
low-temperature optical spectroscopy \cite{Puchkov95,Yoon98}:
\begin{equation}
\label{eq:polaron} \sigma(\omega,T)= \sigma(0,T) \frac{\sinh (4 E_{b}
\hbar\omega / \Delta^2)}{4 E_{b} \hbar\omega / \Delta^2} \,
e^{-(\hbar\omega)^{2}/\Delta^{2}}  \quad
\end{equation}
Here $\sigma(0,T)$ is the dc-conductivity, $E_{b}$ is the SP
binding energy, $\Delta\equiv 2 \sqrt{2 E_{b} E_{\rm vib}}$, and
$E_{\rm vib}$ is the characteristic vibrational energy being
$k_{B}T$ in the high-temperature regime and $\hbar\omega_{\rm
ph}/2$ at low temperatures ($k_B T < \hbar\omega_{\rm ph}$,
$\omega_{\rm ph}$ is a phonon frequency). Eq.~(\ref{eq:polaron})
reproduces the broad resonance with a maximum at $2 E_{b}$ (inset
of Fig.~\ref{fit}). The {\it low-energy} SP absorption is well
represented by the optical conductivity, Eq.~(\ref{eq:polaron}).
However, one would have to consider the consequences of a high
polaron concentration for the absorption in order to understand
the slow decay on the {\it high-frequency} side.

As shown in the inset of Fig.~\ref{fit}, the binding energy $2
E_{b}$ increases towards the metal-insulator transition at $T_{\rm
MI}$, and remains constant above $T_{\rm MI}$ reflecting the
significant role of magnetic interactions in a polaronic system as
well as the importance of electron-phonon interactions in the
formation of the insulating phase. For LSMO the increase of
$E_{0}$ is weaker corresponding to a weaker electron-phonon
interaction. As $T_{\rm MI}$ is above room temperature for LSMO,
$\sigma$ of the insulating phase was not accessible in our
experimental set-up.

\begin{figure}[t]
\centering
\includegraphics[width=.5\textwidth,clip]{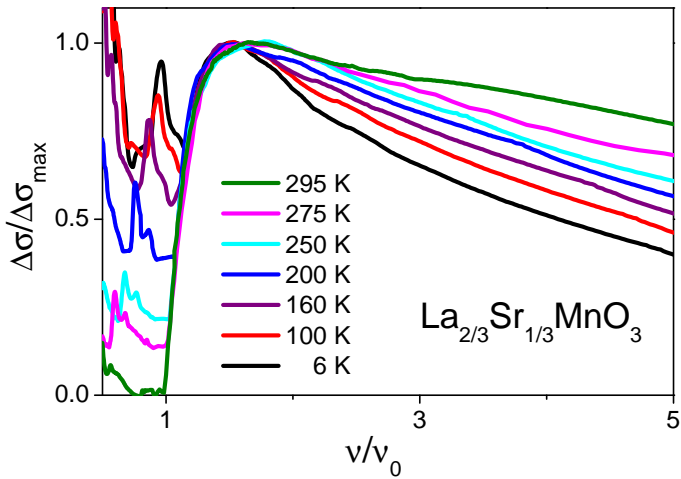}
\caption[]{\label{scaling} Rescaled optical
conductivity of LSMO: $\sigma$ is scaled by the maximum of the 
resonance, $\nu$ is scaled by the threshold value $\nu_0$ of the
onset of the polaronic peak. For $\Delta\sigma$ we subtracted a
constant background $\sigma_{\rm bg}=$370~$\Omega^{-1}$cm$^{-1}$
(cf. Fig.~\ref{fit}). The scaling at the low frequency slope is
nearly independent on the precise value of $\sigma_{\rm bg}$.}
\end{figure}

Concerning LCMO, the SP picture is well accepted and both the
high-temperature dc-resistivity \cite{Worledge96} and the rather
symmetric line shape close to the maximum of the polaronic
excitation suggest that the transport is controlled by incoherent
tunnelling of SP. For LSMO, the interpretation of the polaron peak
in terms of LP is not as obvious. However, an observation related
only to LSMO proves to be most valuable in restricting the
possible scenarios:

If we scale the frequency by the $T$-dependent threshold value
$\nu_0$ and the optical conductivity by its maximum value at the
polaron peak, we find a universal low-energy slope independent of
temperature (see Fig.~\ref{scaling}). This scaling signifies that
the low-energy scattering processes which contribute to the
absorption for this frequency range relative to $\nu_0$ are of the
same origin independent of temperature, related band narrowing,
and other energy scales.

We challenge that Emin's phenomenological approach is applicable
here: (i) it would reproduce the scaling if the dimensionless
polaron radius $\varrho \equiv R \sqrt{2mE_0}/\hbar$ were the same
for all temperature sets; however, $\varrho$ varies from 0.6 to
0.8, a range of values too large to convincingly support the
scaling; (ii) the polaron radius $R/a= 0.8$ is small and local
lattice effects have to be accounted for; (iii) there is no
hole-band minimum available to which the excited charge carriers
could scatter. Furthermore, the energy difference from the
mobility edge below which the trapped polarons would reside to the
hole band minimum would have to be the threshold frequency. This
frequency is about $540$~cm$^{-1}$ at 6~K and is too small for the
addressed energy difference.

\begin{figure}[t]
\centering
\includegraphics[width=.5\textwidth,clip]{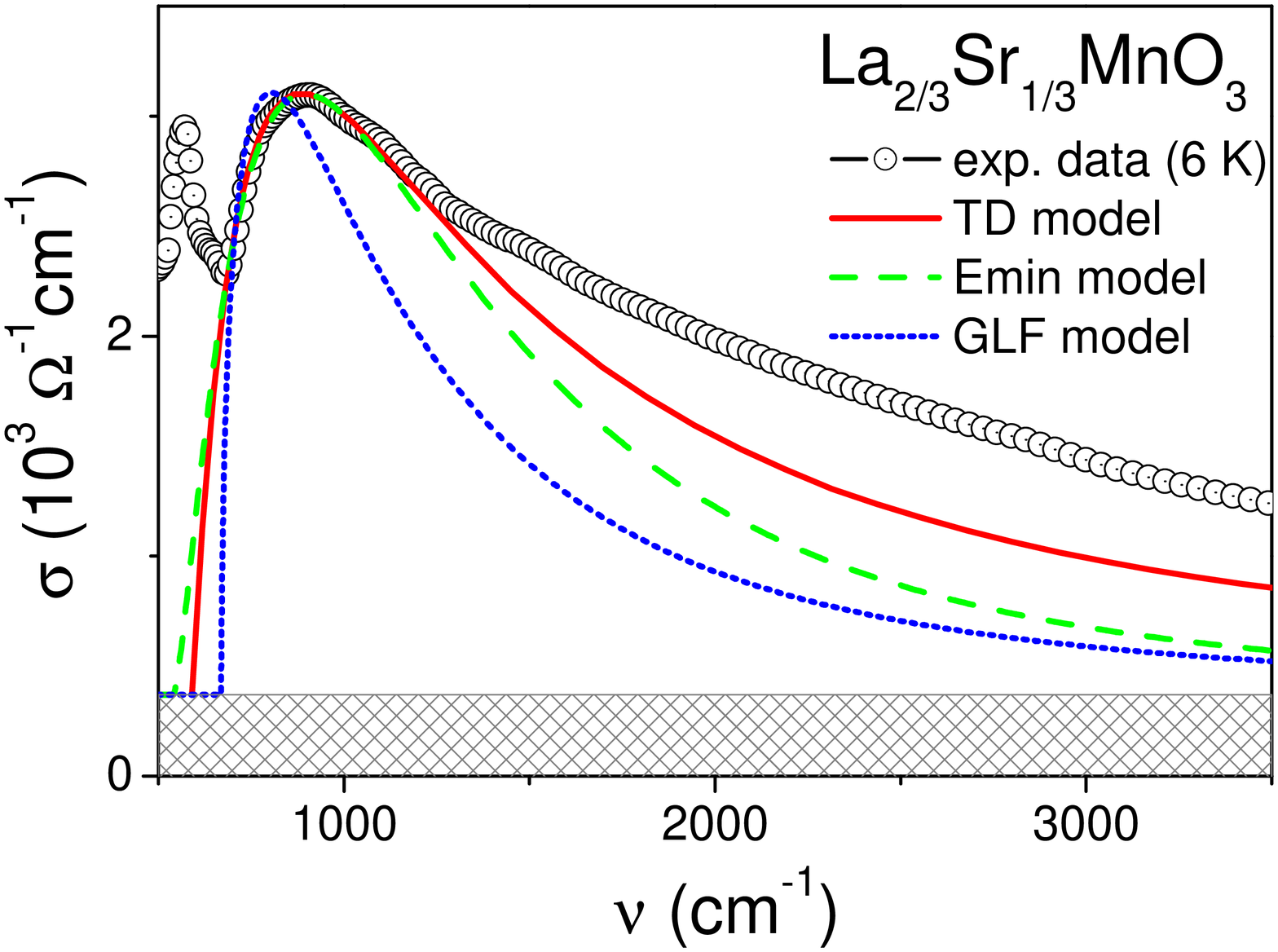}
\caption[]{\label{theory} Comparison of the low temperature MIR
optical conductivity  to $\sigma(\nu)$ from various model
calculations: the solid line refers to the weak coupling approach
of Tempere and Devreese modified for an on-site Hubbard
interaction, well approximated by Eq.~(\ref{eq:ScalingForm}); the
dashed line is the result of the phenomenological approach for
self-trapped large polarons by Emin, Eq.~(\ref{eq:EminPolaron});
the dotted curve is the weak coupling single-polaron result (GLF),
Eq.~(\ref{eq:GLF}).}
\end{figure}

Consequently the modelling has to be revised as to implement LPs with finite
mass and a finite density of charge carriers (Fermi edge). In weak coupling
theory, the generalization of the single polaron absorption by Gurevich, Lang
and Firsov (GLF) \cite{Gurevich62} (dotted curve in Fig.~\ref{theory},
$\alpha$ is the coupling strength)
\begin{equation}
\label{eq:GLF}
\sigma_{\rm GLF}(\omega)\; =\;
   \alpha\; n_p\; \frac{2}{3} \frac{e^2}{m}
       \frac{1}{\omega_0}
  \,(\frac{\omega_0}{\omega})^3
   \;\sqrt{\frac{\omega}{\omega_0} -1}
\end{equation}
to the situation of many-particle absorption was carried through
by Tempere and Devreese for a Coulomb gas \cite{Tempere01}. The
optical conductivity for a similar approach, which builds on the
Hubbard model, is presented by the solid curve in
Fig.~\ref{theory}. For the manganites in the ferromagnetic,
metallic regime, we take the following standard simplifications
for the low-temperature evaluation: (a) only one spin direction
prevails (spinless fermions) due to double exchange and strong
Hunds coupling, (b) there are two degenerate orbital states (the
two $e_g$-levels), (c) the interaction is represented by the local
Hubbard $U$ for two fermions on the same site in two $e_g$ orbital
states. We adjust the parameters as follows: the density of charge
carriers is $n_p =6 \times 10^{21}$~cm$^{-3}$, which is the
stoichiometric number of doped holes and which is in the range of
what has been estimated from Hall measurements \cite{Mandal98}.
Using $m\simeq 3m_e$, consistent with specific heat measurements
\cite{Salamon01}, we then estimate $E_f/h=3270$~cm$^{-1}$. Since
$\nu_0 = 540$~cm$^{-1}$ for the 6~K data, we fix the ratio $h\nu_0
/E_f = 0.17$, i.e.~we are in the adiabatic regime. Finally, we
vary $U/E_f$ in the range $0.1,$~...~$,5$. For values of $h\nu_0
/E_f < 0.3$, the absorption shape is nearly independent on
$U/E_f$, except that the magnitude of $\sigma_{\rm max}$ scales
down with increasing $U/E_f$. For $U/E_f=0.1$, we have
$\sigma_{\rm max}= \alpha \times 2.37 \times 10^3~\Omega^{-1}{\rm
cm}^{-1}$ which should fit the data with $\alpha$ of the order of
1.


For the considered frequency range above the threshold, we are
always sufficiently close to the Fermi edge in order to
approximate the frequency dependence of the imaginary part of the
dielectric response by ($\omega-\omega_0$), the number of excited
low-energy particle-hole pairs. The scaling of the low-frequency
slope (as suggested in Fig.~\ref{scaling}) is implemented through
the increase of the number of  particle-hole pairs with increasing
frequency (relative to the Fermi edge). The decrease at the
high-energy side is controlled by a $1/\omega^3$-factor, most
easily identified in the force-force correlation function form of
the conductivity. Consequently, we expect the optical conductivity
to approach
\begin{equation}
\label{eq:ScalingForm} \sigma(\omega)\;\propto\;
\left(\frac{\omega_0}{\omega}\right)^3\;
(\frac{\omega}{\omega_0}-1) \; .
\end{equation}
Indeed, this formula matches the frequency dependence of the
numerically calculated $\sigma(\omega)$ (solid line in
Fig.~\ref{theory}). The model reproduces the observed shape of the
polaron peak quite convincingly and it accounts for the scaling.
However, it does not provide a mechanism for the observed shift of
the threshold with temperature --- which is probably controlled by
the electron-phonon coupling strength. This necessitates to extend
the considerations to intermediate or strong coupling situations,
first and foremost in the adiabatic limit. Such an evaluation is
not available yet. It has to involve the collective response of
the Fermi sea to the formation of a potential well during the
absorption process which leads to singularities in the optical
conductivity, well known from the the X-ray edge problem. Recoil
of the polaronic lattice deformation, which is taken up in the
excitation process of the charge carrier to the Fermi edge
\cite{Gavoret69,Ruckenstein87}, flattens the singularity and might
be responsible for the observed shape. Though the consequences for
the shape of the absorption spectrum are speculative, the singular
response is supposed to be present in any model with a sudden
creation of a heavy or localized scatterer in the presence of a
Fermi sea, and it will have to be investigated.

Apart from these considerations, the simple modelling presented
above already characterizes the observed polaron excitations
sufficiently well and we are in the position to identify the
nature of the polaronic processes in LSMO and LCMO films as large
and and small polarons, respectively.

\begin{acknowledgments}
We acknowledge discussions with K.-H.~H\"ock, T.~S.~Nunner, R.~Hackl, A.~Nucara
and P.~Calvani. We thank J. Goldfuss and W. Westerburg for placing the samples
at our disposal. The  research was supported by  BMBF (13N6917, 13N6918A) and
partly by DFG via SFB 484 (Augsburg).
\end{acknowledgments}

\newcommand{\noopsort}[1]{}

\end{document}